\documentclass[iop]{emulateapj}
\usepackage{bm}
\usepackage{multirow}
\usepackage{color}
\usepackage{float}
\usepackage{natbib}
\usepackage{amsmath}
\usepackage{amsfonts}
\usepackage{amssymb}
\usepackage{epsfig}
\usepackage{hyperref}
\usepackage{graphicx}
\usepackage{booktabs}

\newcommand{\beq}{\begin{equation}}
\newcommand{\eeq}{\end{equation}}
\newcommand{\beqa}{\begin{eqnarray}}
\newcommand{\eeqa}{\end{eqnarray}}

\begin{document}

\title{Multi-wavelength lens reconstruction of a \textit{Planck} \& \textit{Herschel}-detected star-bursting galaxy}
\author{Nicholas Timmons$^{1}$, Asantha Cooray$^{1}$, Dominik A. Riechers$^{2}$, Hooshang Nayyeri$^{1}$, Hai Fu$^{3}$, Eric Jullo$^{4}$, Michael D. Gladders$^{5}$, Maarten Baes$^{6}$, R. Shane Bussmann$^{2}$, Jae Calanog$^{7}$, David L. Clements$^{8}$, Elisabete da Cunha$^{9}$, Simon Dye$^{10}$, Stephen A. Eales$^{11}$, Cristina Furlanetto$^{10,12}$, Joaquin Gonzalez-Nuevo$^{13}$, Joshua Greenslade$^{8}$, Mark Gurwell$^{14}$, Hugo Messias$^{15}$, Micha\l{} J. Micha\l{}owski$^{16}$, Iv\'{a}n Oteo$^{16,17}$, Ismael P\'{e}rez-Fournon$^{18,19}$, Douglas Scott$^{20}$, Elisabetta Valiante$^{11}$}

\affiliation{$^{1}$Department of Physics  and  Astronomy, University  of California, Irvine, CA 92697}
\affiliation{$^{2}$Department of Astronomy, Cornell University, Ithaca, NY 14853, USA}
\affiliation{$^{3}$Department of Physics \& Astronomy, University of Iowa, Iowa City, IA 52242}
\affiliation{$^{4}$Aix-Marseille Universit\'{e}, CNRS, LAM (Laboratoire d'Astrophysique de Marseille) UMR 7326, 38 rue Joliot-Curie, 13388 Marseille Cedex, France}
\affiliation{$^{5}$The Department of Astronomy and Astrophysics, and the Kavli Institute for Cosmological Physics, The University of Chicago, 5640 South Ellis Avenue, Chicago, IL 60637, USA}
\affiliation{$^{6}$Sterrenkundig Observatorium, Universiteit Gent, Krijgslaan 281S9, 9000 Gent, Belgium}
\affiliation{$^{7}$Department of Physical Sciences, San Diego Miramar College, San Diego CA, 92126}
\affiliation{$^{8}$Physics Department, Blackett Lab, Imperial College, Prince Consort Road, London SW7 2AZ, UK} 
\affiliation{$^{9}$Centre for Astrophysics and Supercomputing, Swinburne University of Technology, Hawthorn, Victoria 3122, Australia}
\affiliation{$^{10}$School of Physics and Astronomy, The University of Nottingham, University Park, Nottingham, NG7 2RD, UK  }
\affiliation{$^{11}$School of Physics and Astronomy, Cardiff University, Queens Buildings, The Parade, Cardiff CF24 3AA, UK}
\affiliation{$^{12}$CAPES Foundation, Ministry of Education of Brazil, Bras\'ilia/DF, 70040-020, Brazil}
\affiliation{$^{13}$Departamento de F\i{}sica, Universidad de Oviedo, C. Calvo Sotelo s/n, 33007 Oviedo, Spain}
\affiliation{$^{14}$Harvard-Smithsonian Center for Astrophysics, 60 Garden Street, Cambridge, MA 02138}
\affiliation{$^{15}$Instituto de Astrof\'isica e Ci\^encias do Espa\c{c}o, Universidade de Lisboa, OAL, Tapada da Ajuda, PT 1349-018 Lisboa, Portugal}
\affiliation{$^{16}$Institute for Astronomy, Royal Observatory, Blackford Hill, Edinburgh, EH9 3HJ, United Kingdom}
\affiliation{$^{17}$European Southern Observatory, Karl-Schwarzschild-Str. 2, 85748 Garching, Germany}
\affiliation{$^{18}$Instituto de Astrof\'{\i}sica de Canarias (IAC), E-38200 La Laguna, Tenerife, Spain}
\affiliation{$^{19}$Departimento de Astrof\'{\i}sica, Universidad de La Laguna, E-38206, La Laguna, Tenerife, Spain}
\affiliation{$^{20}$Department of Physics \& Astronomy, University of British Columbia, 6224 Agricultural Road, Vancouver, BC V6T 1Z1, Canada}

\begin{abstract}

We present a source-plane reconstruction of a {\it Herschel} and {\it Planck}-detected gravitationally-lensed dusty star-forming galaxy (DSFG) at $z=1.68$ using
{\it Hubble}, Sub-millimeter Array (SMA), and Keck observations. The background sub-millimeter galaxy (SMG) is
strongly lensed by a foreground galaxy cluster at $z=0.997$ and appears as an arc of length $\sim 15^{\prime \prime}$ in the optical images. The continuum dust emission, as seen by SMA,
is limited to a single knot within this arc. We present a lens model with source plane reconstructions at several wavelengths to show the difference in magnification between the stars and dust, and highlight the importance of a multi-wavelength lens models for studies involving lensed DSFGs. 
We estimate the physical properties of the galaxy by fitting the flux densities to model SEDs leading to a magnification-corrected star formation rate of $390 \pm 60$ 
M$_{\odot}$ yr$^{-1}$  and a stellar mass of $1.1 \pm 0.4\times 10^{11}$ M$_{\odot}$. These values are consistent with high-redshift massive galaxies that
have formed most of their stars already. The estimated gas-to-baryon fraction,
molecular gas surface density, and SFR surface density  have values of $0.43 \pm 0.13$, $350 \pm 200$ M$_{\odot}$ pc$^{-2}$,  and $\sim 12 \pm 7~$M$_{\odot}$ yr$^{-1}$ kpc$^{-2}$,
respectively. The  ratio of star formation rate surface density to molecular gas surface density puts this among the most star-forming systems, similar to other measured SMGs 
and local ULIRGS.
\end{abstract}

\maketitle

\section{Introduction}

\begin{figure*}[t]
\begin{center}
 \includegraphics[scale=1.2]{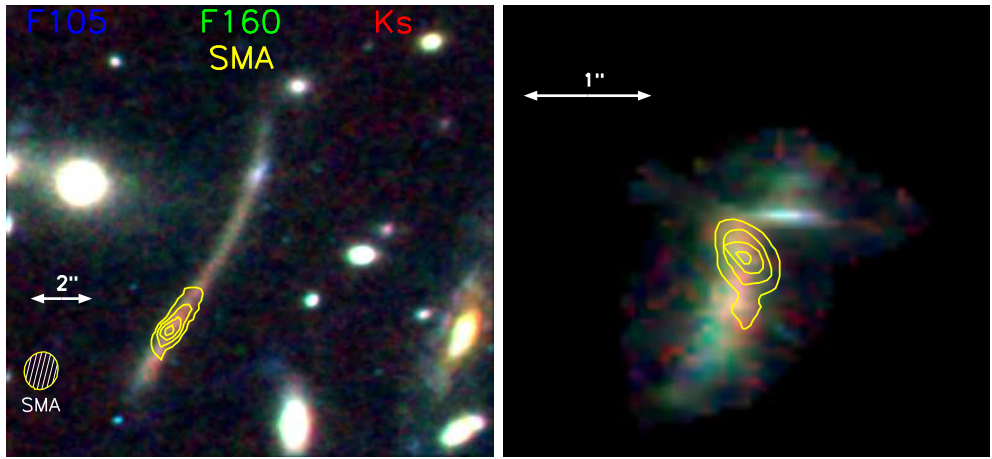}
  \caption {{\it Left:} Three color image of HATLAS J132427+284452 using {\it Hubble}/WFC3 F105W (blue) and F160W (green), and
Keck NIRC2 $K_{\rm s}$ (red) bands with Submillimeter Array (SMA) $870~\mu$m band emission contours are overlaid. The SMA contours are at $3\sigma$, $6\sigma$, $9\sigma$ and $12\sigma$, where $\sigma$ is the rms noise ($0.6$ mJy beam$^{-1}$). 
The dust emission, and thus the {\it Herschel} and primary {\it Planck} source,
associated with the DSFG is concentrated in the area of the yellow contours while the optical emission extends over an arc of
$ \sim 15^{\prime \prime}$. For reference, we show the SMA beam in the bottom left.
{\it Right:} Three color source plane reconstruction with SMA source plane contours overlaid with the same contouring steps as left (see Section 4 for the lens
reconstruction). The spatial resolution of the reconstruction is $\sim 0.06~ ^{\prime \prime}$ pixel$^{-1}$ or $\sim 0.5~$kpc pixel$^{-1}$.}
\end{center}
\end{figure*}

\begin{figure}
 \includegraphics[scale=.22]{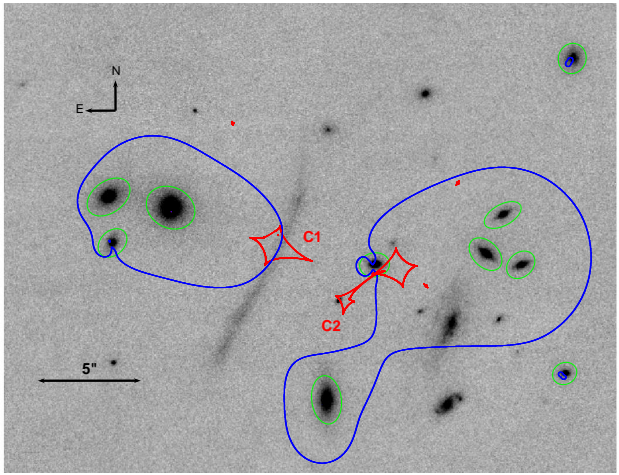}
  \caption {Keck/NIRC2 $K_{\rm s}$-band image with the critical and caustic (C1 and C2) lines over-plotted in blue and red, respectively. Circled in green are the foreground lens galaxies used in constructing the lens model. In addition to individual galaxies the lensing reconstruction requires extended potential associated with the two galaxy groups/clusters to the east and west of the lensing arc. 
}
\end{figure}

In recent years, large area far-infrared and sub-millimeter surveys, for example, the \textit{Herschel}-Astrophysical TeraHertz Large Area Survey (H-ATLAS) \citep{Eales2010}, have 
allowed the efficient selection of gravitationally lensed high-$z$ dusty star-forming galaxies (DSFGs; e.g., \citealt{Negrello2007,Negrello2010,Gonzalez2012,Wardlow2013,Bussmann2013,Nayyeri2016}). These DSFGs (see \citealt{Casey2014} for a recent review) have star formation rates (SFRs) of $\sim 10^2$--$10^3~ $M$_{\odot}$ yr$^{-1}$, with typical stellar mass of $\sim 10^{11}$--$10^{12}~$M$_{\odot}$, and are generally found during the peak epoch of galaxy formation and evolution at $z \sim 1-4$. Such rapid star- formation has a short lifetime ($< 0.1$ Gyr) and is rare in the local Universe \citep{Tacc2010}. Luminous and ultra-luminous infrared galaxies (LIRGs and ULIRGS), of which DSFGs are an analog, contribute significantly ($\sim 70\%$) to the cosmic star formation at $z = 1$ \citep{Lefloc2005}. Recent studies have shown that DSFGs may differ from ULIRGs in that their star-forming regions may be more spatially extended (e.g., \citealt{Younger2008,Ivison2011,Riechers2011}). There is evidence to suggest DSFGs are likely an early stage of today's massive elliptical galaxies (e.g., \citealt{Lilly1999,Swinbank2006,Lapi2011,Fu2013}). DSFGs are usually faint at rest-frame optical wavelengths due to dust obscuration, but are bright in the rest-frame far-IR, making sub-mm surveys the perfect tool to study DSFGs \citep{Negrello2010}.

While wide area surveys with {\it Herschel} and ground-based instruments have increased the sample sizes of DSFGs at sub-mm wavelengths, due to
limitations associated with existing instruments in sensitivity and spatial resolution,
our ability to conduct detailed investigations on the physical properties of DSFGs has been severely hampered.
Thankfully, strong gravitational lensing can be used to overcome these limitations. The flux amplification as a result of gravitational lensing allows for the detection of otherwise intrinsically fainter dust obscured galaxies and the associated spatial enhancement allows spatially resolved imaging observations with existing facilities (e.g., \citealt{Fu2012,Messias2014}).

H-ATLAS J132427.0+284452 (hereafter HATLAS J132427) peaks at $350~ \mu$m with a flux density of $\sim 380 \pm 8$ mJy (from \textit{Herschel} Spectral and Photometric Imaging Receiver, SPIRE). It is also identified 
in the all-sky maps from {\it Planck} \citep{Plank2011} as PLCKERC857 G047.32+82.53 ($1.3 \pm 0.15$ Jy) at 857 GHz (350 $\mu$m) in the {\it Planck} Early Release Compact Source Catalog (ERCSC; \citealt{Plank2011}). Although the {\it Planck} detected flux density is $\sim 4\times$ larger than {\it Herschel}/SPIRE measurement in H-ATLAS, 
the difference can be explained as due to the large 3-5 arcmin beam of {\it Planck} measurements which may cause blending in an over-dense field. Such a difference is also present in a
previous {\it Planck}-detected H-ATLAS lensed source. H-ATLAS J114637.9-001132 \citep{Fu2012} is detected by {\it Planck} with a flux density of S$_{350} = 2.1 \pm 0.8~$Jy but in \textit{Herschel} the flux density is measured to be S$_{350} = 378 \pm 28~$mJy corresponding to a $\sim 5\times$ larger {\it Planck} flux density much like HATLAS J132427.
Despite the {\it Planck} flux being uncertain the detection is validated through other 
observations and confirms {\it Planck}'s ability to detect the brightest lensed DSFGs (see \citealt{Canameras2015}).

In this paper we present new \emph{HST}, SCUBA2 and Keck observations of HATLAS J132427 along with previous multi-wavelength observations to create a complete profile of this {\it Planck} and {\it Herschel}-detected DSFG. In Section~2 we describe the observations and data reduction procedures. In Section~3 we describe previous and archival observations used in the analysis. In Section~4 we use high resolution imaging to construct a lens model and calculate the magnification factors. In Section~5 we model the spectral energy distribution (SED) and derive physical properties from the fit. In Section~6 we discuss the derived properties of HATLAS J132427 and compare them to other SMGs and DSFGs. We conclude with a summary in Section~7. Throughout we make use of the standard flat-$\Lambda$CDM cosmological model with $H_0$= 70 km s$^{-1}$ Mpc$^{-1}$ and $\Omega_{\Lambda}$=0.73.

\section{Observations}

Early observations of HATLAS J132427 are presented in \citet{George2013}. Here we present
new Keck, SCUBA2, {\it Hubble}/WFC3 imaging data and {\it Hubble}/WFC3 grism observations. Figure 1 shows a three color image of the source using \emph{HST} (F105W and F160W bands) and Keck ($K_{\rm s}$ band) imaging along with SMA contours overlaid to show the spatial variations of the source at different wavelengths. Figure 2 shows Keck NIRC2 $K_{\rm s}$-band imaging with the critical and caustic lines used in the lens model. 

\begin{table}
\caption{Observed Properties} 
\centering
\resizebox{0.75\columnwidth}{!}{%
\begin{tabular}{c c} 
\hline\hline\\ [0.1ex] 
Parameter &  Value\\ [0.5ex]
\hline\\ [0.5ex]
R.A., DEC & 13:24:27.206 +28:44:49.40\\[0.5ex] 
$z_{\rm source}$ & $1.676 \pm 0.001$\\[0.5ex]
$z_{\rm lens}$  & $0.997 \pm 0.017$\\[0.5ex]
\hline\hline\\[0.5ex]
\end{tabular}
}
\label{table:PHOT1} 
\end{table}

\subsection{Keck/NIRC2}
We obtained a 1680 second exposure in $H$ band with an airmass of 1.02 and a 3840 second exposure in $K_{\rm s}$ band with an airmass of 1.36 (PI: Cooray) on 4 February 2012 with the KeckII/NIRC2
instrument aided with the laser guide-star adaptive optics system (LGSAO; \citealt{Wizinowich2006}). The imaging observations made use of a pixel scale at
$0.04''~$pixel$^{-1}$ for both filters. Custom IDL scripts were used to reduce the data following the procedures in \citet{Fu2012,Fu2013} which includes a dark subtraction, bad pixel masking, background subtraction as well as flat-fielding. The $K_{\rm s}$-band image was flux calibrated using UKIDSS \citep{Lawrence2007} $K$-band photometry. The $H$-band 
image was flux calibrated using a common set of bright stars detected in NIRC2 image and in the {\it Hubble}/WFC3 F160W band image.

\subsection{Hubble/WFC3}

\begin{figure*}[t]
\begin{center}
 \includegraphics[scale=.85]{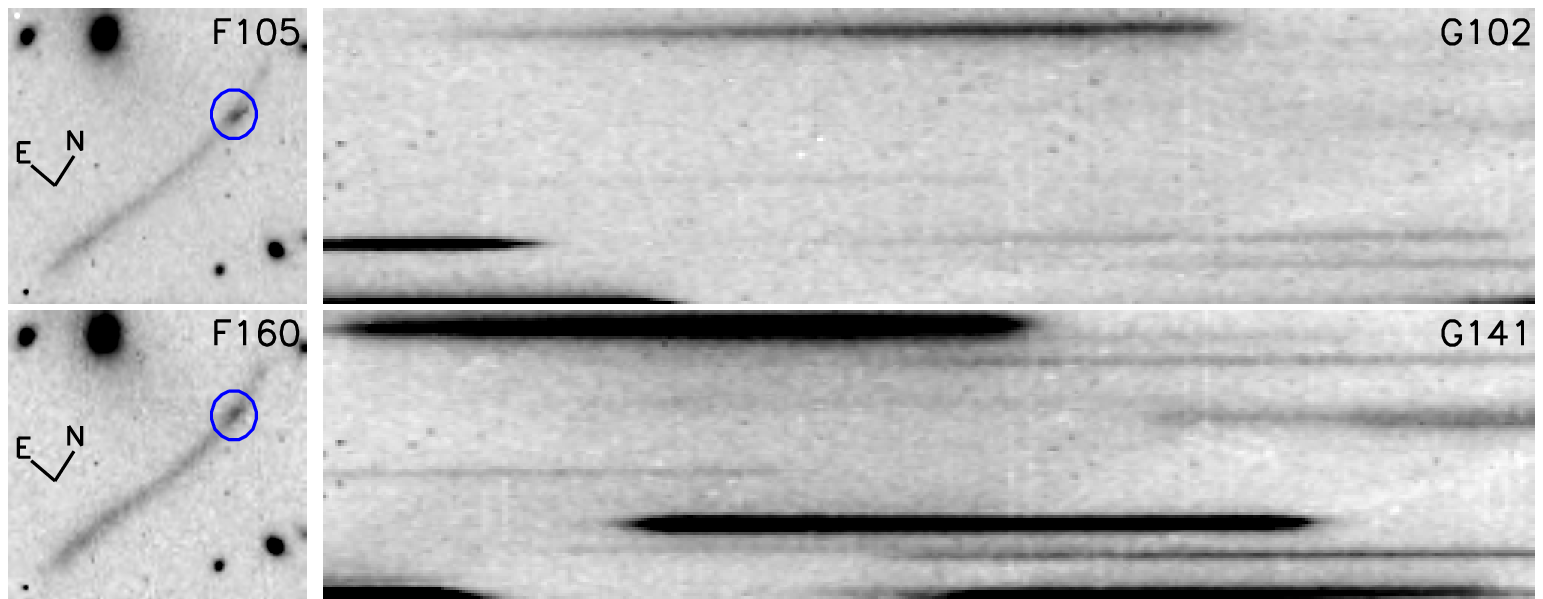}
 \includegraphics[scale=.45]{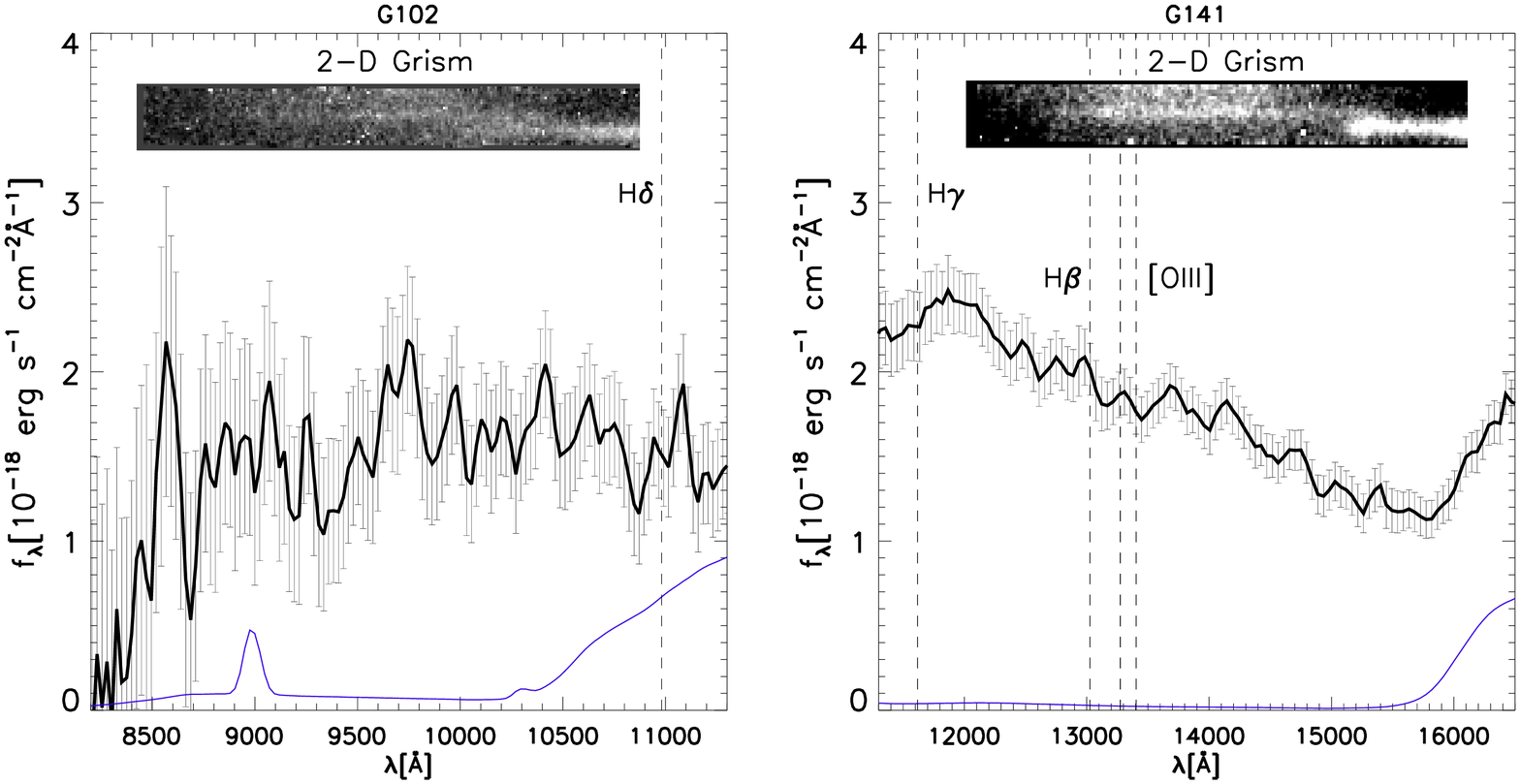}
  \caption {{\it Top Left:} The direct image in each of the WFC3 imaging filters oriented so that the dispersion direction of
the grism is horizontal. {\it Top Right:} The two-dimensional grism images
of HATLAS J132427. The top panel shows G102 and F105W
images while the middle panel shows G141 and F160W
images. {\it Bottom:} The extracted 1D spectra from the G102 and G141 slit-less spectra. The blue line is the estimated contamination coming from other spectra in the field. The 2D grism stamps are inlayed with vertical lines corresponding to useful emission lines over-plotted. Despite the presence of continuum emission no emission lines were detected. Due to contamination from other spectra in the field only the bright northern clump which has been circled in blue had an extractable continuum. It was not possible to extract a 1D spectrum from the southern clump associated with the radio detection.
}
\end{center}
\end{figure*}

{\it Hubble}/WFC3 observations of HATLAS J132427 were completed with three orbits under GO program 13399 in Cycle 21 (PI: Cooray). We obtained a total of ten exposures including two direct images (F105W and F160W) and eight grism observations. The F105W observation had a total exposure time of 453 seconds while the F160W observation had a total exposure time of 353 seconds. Six of the grism observations were taken with the G102 (800 nm--1150 nm) grism for a total exposure time of 5218 seconds. The remaining two grism observations were taken with the G141 (1075 nm--1700 nm) grism for a total exposure time of 2406 seconds.

We made use of the calibrated \emph{HST} imaging and grism data from the CALWF3 reduction pipeline, as provided by the Space Telescope Science Institute \footnote{www.stsci.edu/hst/wfc3/pipeline/wfc3\_pipeline}. The spectra for individual objects in the image were extracted with the aXe software package \citep{Kummel2009}. The data products include the two-dimensional combined grism stamp for each object as well as flux-calibrated one-dimensional spectra, contamination estimates, and error estimates. Similar analysis and reduction steps for the other target, (HATLASJ1429-0028) in GO program 13399 in Cycle 21 are described in \citet{Timmons2015}.

The top portion of Figure 3 shows the direct imaging for the F105W and F160W filters aligned so that the dispersion direction of the grism is horizontal. Figure 3 also shows the two-dimensional stamps for the two grism filters. The bottom portion of Figure 3 shows the extracted one-dimensional spectra for each grism filter with a close up view of the two-dimensional continuum shown as an inset. The 2D stamp and the 1D spectra come from the bright northern clump as can be seen in Figure 3. Only the northern clump had a detectable continuum that was not overly contaminated by other spectra in the field. This clump has been circled in blue in the F105W and F160W images in Figure 3. The expected emission lines at $z = 1.68$ are shown in the 1D spectra of Figure 3 and it is clear there is no significant line detection in either of the grism spectra, and so we cannot conduct line ratio diagnostics on HATLAS J132427. This is due to the low surface brightness of the galaxy compared to the source detected in \citet{Timmons2015} which involved bright multiply-imaged star-forming knots. Unfortunately, due to the overlapping grism spectra from nearby galaxies, we cannot integrate longer to improve the signal-to-noise of the spectrum from our target.

\subsection{SCUBA2}

This source, and the field around it, was observed by the SCUBA2 bolometer array camera on the JCMT \citep{Holland2013}. These observations were part of a broader program following up sources in the H-ATLAS survey (M13AU12, PI D.L. Clements). The observations of the field around HATLAS J132427 were made on 2013 April 8th and 12th using the standard pseudo-circular DAISY observing sequence for small and compact sources. This provides maps of a circular region of roughly 350 arcseconds in radius around the target position. The integration time in this field is a function of position, with the central regions receiving greater integration time than the outer regions. Five separate DAISY maps of HATLAS J132427 were made, three on April 8th, two on April 12th. The conditions for these observations were rated grade 3, indicating $\tau_{\rm 225GHz}$ 0.08 to 0.12. These conditions are adequate for 850~$\mu$m observations but not for good 450~$\mu$m photometry.

The data were reduced in the standard manner using the SMURF software provided by the observatory. The SMURF iterative mapmaker \textit{makemap} produced individual maps for each of the five subintegrations using the reduction recipe optimized for blank fields with corrections for atmospheric opacity. The five resulting maps were then combined using the mosaic tool to produce a final image which was then match filtered to optimize the S/N of unresolved sources. The final image was then trimmed to produce a 350 arcsecond radius field. The final map has a total integration time of 1850 s at its center, where HATLAS J132427 is located, falling to 450 s at the edges. HATLAS J132427 is detected at the centre of the final images with an 850~$\mu$m S/N ratio of $\sim$ 30 and a flux of 43 $\pm$ 1.2 mJy. It is interesting to note that five other 850~$\mu$m sources are detected at $>4 \sigma$ in the final map, suggesting the presence of a moderate over-density of sub-mm sources around HATLAS J132427.

\section{Previous and Archival Observations}

\begin{figure}
 \includegraphics[trim= 3cm 0 0 0,clip,scale=.33]{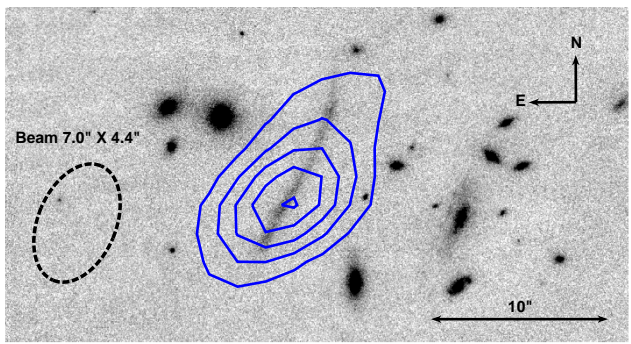}
  \caption {Keck/NIRC2 $K_{\rm s}$-band image with CARMA contours overlaid. The contours are at $2\sigma$, $4\sigma$, $6\sigma$, $8\sigma$ and $10\sigma$, where $\sigma$ is the rms noise ($0.76$ mJy beam$^{-1}$). For reference, we show the CARMA beam size and orientation.} 
\end{figure}

HATLAS J132427 was first reported as a candidate strongly lensed giant arc at optical wavelengths in \citet{Gladders2003} and its discovery and follow-up as a bright source in Herschel data is discussed in \citet{George2013}. The following is a summary of previous or archival observations that were used for the present analysis. The flux densities are shown in Table 2.

\textit{Herschel} Photoconductor Array Camera and Spectrometer (PACS) \citep{Poglitsch2010} data at 100 $\mu$m and 160 $\mu$m were collected as a part of the OT1 program (OT\_RIVISON\_1). The total integration time of 360 s reaching $\sigma \sim$ 10 mJy for 100 $\mu$m and $\sigma \sim$ 12 mJy for 160 $\mu$m. \textit{Herschel}/SPIRE Fourier Transform Spectrometer (FTS) \citep{Griffin2010} observations were completed on 2 August 2012. The wavelength coverage was $\lambda_{\rm obs} = 194-671~ \mu$m and the total observing time was 3.8 h. The data resulted in the discovery of
the bright [CII]/158 $\mu$m emission line with a peak flux density of $\sim 0.8~$Jy, allowing the redshift  of $z=1.68$ to be measured directly, for the first time, from far-infrared spectroscopy. While the PACS data are used for the SED analysis the FTS spectrum is not. It is shown in Figure 6 but is not used in the SED analysis due to the presence of the 
bright [CII]158 $\mu$m emission line. 

As a part of program 2011B-S044, 870 $\mu$m imaging data were taken with the Submillimeter Array (SMA) (PI: Bussmann). The total integration time of 9.7 h was taken in the compact, extended, and very extended array configurations, with baselines of 20-400 m. 1924-292, a blazar, was utilized as a bandpass calibrator and Titan was used for the flux calibration \citep{Bussmann2013}. The effective beam size is $1.66''$ and the rms is $6$ mJy beam$^{-1}$. The SMA continuum is shown in Figure 1 and is used in the lensing model.

The CO J= $2 \to 1$ line ($\nu_{\rm rest} = 230.538$ GHz, $\nu_{\rm obs} = 86.0$ GHz at $z=1.68$) was detected by the Combined Array for Research in Millimeter-wave Astronomy (CARMA; PI: Riechers). The observations were conducted on 23 November 2012 using the D configuration ($11$-$146~$ m baselines). The beam size was $7'' \times 4.4''$ and a rms noise of $0.76$ mJy beam$^{-1}$. The total on-source time was 2.3 h while two blazars 1310+323 and 0927+390 were used to derive the bandpass shape and for complex gain calibration. Figure 4 shows CARMA contours overlaid on the Keck/NIRC2 $K_{\rm s}$-band image.

The Canada-Hawaii-France Telescope (CFHT) was used to image HATLAS J132427 in both the z ($925~$nm) and r ($640~$nm) bands (PI: Yee). The integration times for the $r$ and $z$ bands were 900 s and 600 respectively and the observations were carried out on 5 July 1999 \citep{Gladders2003}. These observations were used to measure $z=0.9~$ for the foreground cluster.

The Institut de Radioastronomie Millimetrique Plateau de Bure Interferometer (IRAM PdBI) was used to obtain 1.1 h of on-source time during November 2012 using six 15 m antennas with the D configuration. The frequency was set to 129.028 GHz. The CO $J = 3 \to 2$ line was detected at $3\sigma$.The flux density measurement is used in the SED analysis.

 HATLAS J132427 is detected by the Wide Field Infrared Survey Explorer (WISE) \citep{Wright2010} in four bands ranging from $3.35 \to 22.09~ \mu$m all used in the SED analysis (Table~2).

\begin{figure*}
\begin{center}
 \includegraphics[scale=.75]{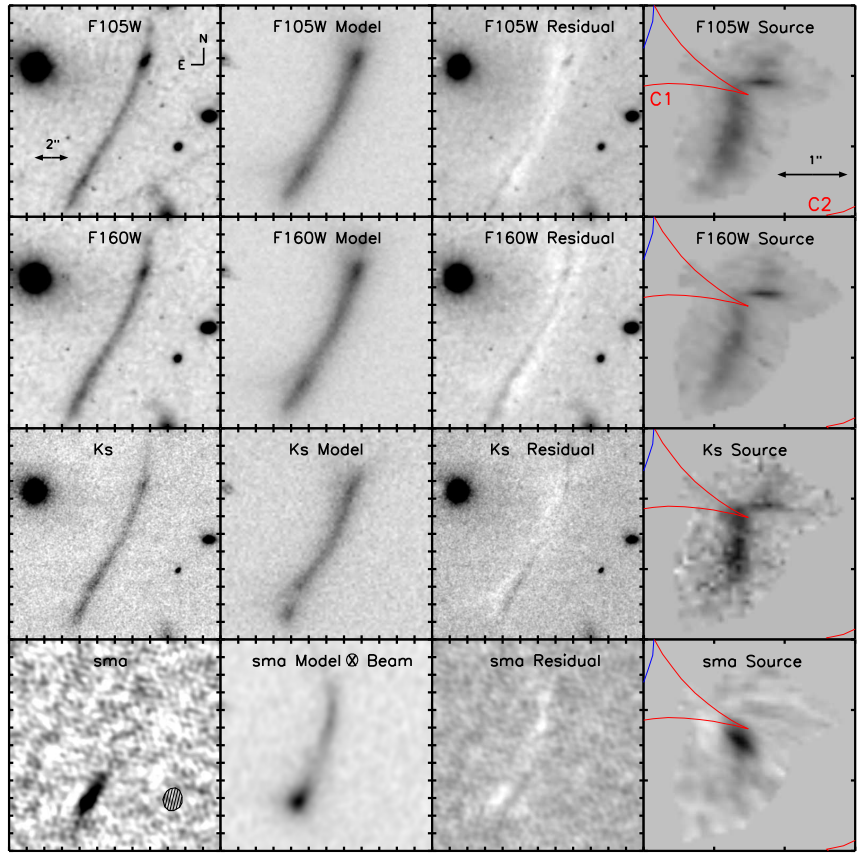}
  \caption {Lens modeling of HATLAS J132427 at several optical/infrared wavelengths and at 870 $\mu$m. {\it 1st column:} The original imaging for the two \emph{HST} bands, as well as Keck and SMA. The beam size and orientation is overlaid on the SMA frame. {\it 2nd column:} Image obtained with the lens model for each band. The SMA model is convolved with a 2D Gaussian model of the SMA beam. {\it 3rd column:} The residual obtained by subtracting the model from the original image. The scale is set to see the areas of over and under subtraction. {\it 4th column:} The source plane reconstruction for each band, with the critical and caustic lines overlaid in blue and red, respectively. The C2 and C2 refer to the caustic lines as shown in Figure 2.}
\end{center}
\end{figure*}

\section{Lens Model}

We make use of the program {\sc LENSTOOL} \citep{Kneib1996,Jullo2007} to reconstruct the lensed galaxy and to derive the magnification factors of HATLAS J132427. Using the \emph{HST} F160W high resolution imaging data, the gravitational potentials contributing to this model are identified using SExtractor \citep{Bertin1996} with their parameters being optimized by the Bayesian Markov chain Monte Carlo (MCMC) sampler used in {\sc LENSTOOL}. For each image (F160W, F105W, $K_{\rm s}$, SMA) the whole arc is broken down into four ellipses of varying size and brightness which are created using measured elliptical sizes and flux densities from SExtractor. These ellipses are then passed through the {\sc LENSTOOL} model to reconstruct the source plane image.

Figure 2 shows Keck/NIRC2 $K_{\rm s}$-band image, with the gravitational potentials used in the model circled in green, and the critical and caustic lines overlaid. From \citet{Gladders2005} the cluster members used in the model are at photometric $z = 0.997 \pm 0.017$ based on $r$ and $z$ band imaging. We assume a constant mass-to-light ratio and adopt a $0.5''$ uncertainty in the position of the critical lines which, corresponds to the thinnest part of the arc and should account for line-of-sight perturbations. The lens galaxies are modeled using a pseudo-elliptical isothermal mass density profile (PIEMD) \citep{Kneib1996}. To create the model the other sources in the field of unknown redshift were also placed at $z = 0.997$. There are a total of 26 galaxies used in the model, most of which are out of view in the figure and do not contribute significantly to the modeled potential. The cluster members which contribute the largest potential to the model are the galaxies which fall in the blue critical lines in the figure.

Compared with observations, models with multiply-imaged systems resulted in lenses that were too large and, thus, unrealistic. Therefore, a model assuming a singly imaged source was utilized. As a main constraint, we assumed that the central thin part of the arc was overlapping the critical line, as has been observed for some very elongated arcs (see the Clone arc in \citealt{Jones2010}). Placing the critical line closer to the arc results in increased stretching. The arc of HATLAS J132427 is very stretched, thus the critical line must overlap with the arc. However, the critical line cannot cross the arc, otherwise there would be two images.

Figure 5 shows the imaging for four bands F105W, F160W, $K_{\rm s}$ and SMA along with their model in the image plane, the residual and the source plane reconstruction. The long arc is detected in the near-IR bands, with the SMA flux only being detected above $3\sigma$ in the southern portion of the arc. The stellar portion of HATLAS J132427 corresponds to the large extended arc suggesting it has a higher magnification than the dust portion. The third column of Figure 5 shows the residual after subtracting the model from the image. It is clear that the model that {\sc LENSTOOL} constructs does not perfectly describe the morphology and does leave residuals. Considering the lack of additional constraints to improve the overall lens model, resulting from a singly-imaged source, we accepted that the current model is likely the best we can presently construct.

 The best fit model gives $\mu_{\rm dust} = 4.9 \pm 1.8$ while the stellar magnification making up the extended arc is $\mu_{\rm stars} = 15.7 \pm 4.3$. In \citet{George2013} a magnification estimate for the molecular gas is derived following \citet{Harris2012} and \citet{Bothwell2013}. Using the J$ = 1 \to 0$ luminosity and the FWHM, $\mu_{\rm Gas}$ is found to be $\sim 11$. Due to the large uncertainty in the FWHM of the gas (e.g., $640 \pm 270$ km s$^{-1}$) the final estimate of the error for the derived value is $ \pm~7$ which is consistent with the magnification values found with the lens model used here. In \citet{Bussmann2013} a lens model for SMA using two galaxies instead of the two cluster components resulted in a magnification of $2.8 \pm 0.4$. The SMA data having just one image in \cite{Bussmann2013} made the model more difficult to constrain whereas the multi-wavelength model presented here includes SMA, Keck, \emph{HST} etc. and can be considered a more complete model of the dust magnification.     

\section{Spectral energy distribution modeling}

\begin{figure*}[t]
\begin{center}
 \includegraphics[scale=.85]{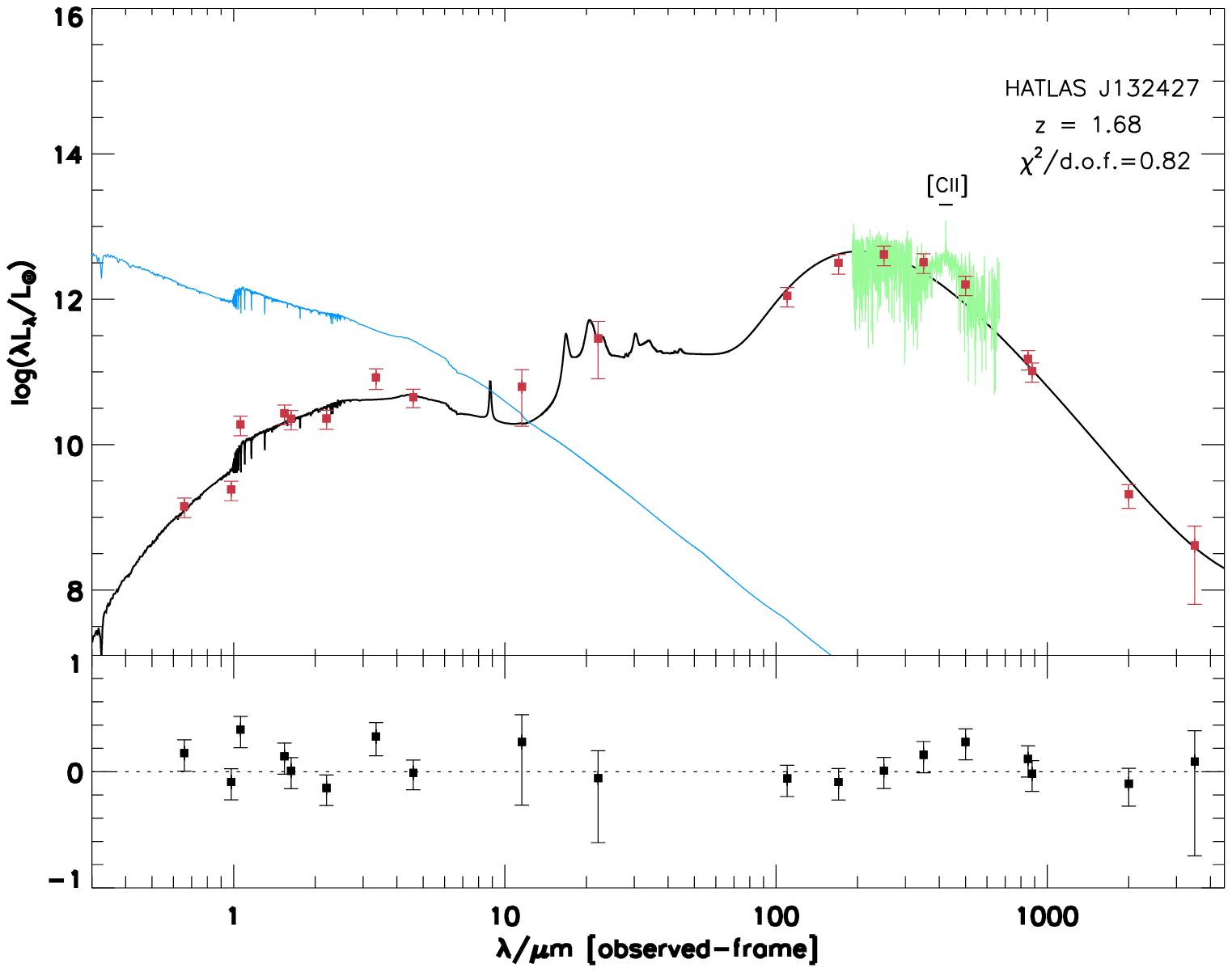}
  \caption {Top: The best-fit SED model is plotted in black while the intrinsic model without dust extinction is plotted in blue. The flux values have been de-magnified based on wavelength. The \textit{Herschel} FTS spectrum is shown in green. The FTS spectrum is not used in the SED fit but is shown here for reference. Bottom: The residuals for each fit.}
\end{center}
\end{figure*}

The spectral energy distribution (SED) of HATLAS J132427 was analyzed using the Multi-wavelength Analysis of Galaxy Physical Properties ({\sc MAGPHYS}) software \citep{daCunha2008}. The {\sc MAGPHYS} package compares the observed flux density values to a library of model SEDs at the same redshift. 
Here we use the new HIGHZ model library of {\sc MAGPHYS} SEDs, which was developed to interpret observations of SMGs from the ALESS survey \citep{daCunha2015}, and should be more appropriate to fit the SEDs of DSFGs at high redshift. 

The photometry for CFHT, \emph{HST} and Keck were done using the SExtractor package \citep{Bertin1996} using a flexible elliptical aperture to account for the elongated nature of the source. The WISE photometry comes from the online WISE catalogs. The remaining photometry comes from \citet{George2013} and is discussed in Section 3. Table 2 lists the observed photometry used in the model fit with a spectroscopic redshift of 1.68. Because there is differential magnification for the dust and stellar components\citep{Calanog2014} the observed fluxes were de-magnified based on wavelength. The stellar fluxes corresponding to the full arc in the SED were de-magnified by $15.7 \pm 4.3$ while the dust portion centered on the lower bright clump was de-magnified by $4.9 \pm 1.8$.

The WISE W3 and W4 bands, at 12 and 22 $\mu$m respectively, posed a problem as the {\sc MAGPHYS} model SED showed those flux densities
were a combination of both stellar and dust emission. In order to account for the uncertainty the error bars were extended to cover the entire magnification range,
with flux densities corrected by 10, corresponding to the average of the dust and stellar magnification factors. Several fits were performed using a lower magnification for the W4 band, corresponding to more dust emission, as well as a higher magnification for the W3 band, corresponding to higher stellar contribution, in the end the average value provided the best-fit.

We note that the SMA flux measurements might be underestimated in the fit due to the short baseline coverage of the observations which could account for the difference between the $870~\mu$m and $850~\mu$m flux values. This could lead to an underestimate of the SFR which is correlated with the total dust luminosity \citep{Kennicutt1998}. The dust temperature is also correlated with the dust luminosity \citep{Chapman2005} and therefore could also be underestimated. The compact configuration of SMA is expected to give an angular resolution of about $9''$ and, considering the large uncertainty and narrow width of the feature, the total flux from SMA should not be resolved out.

Figure 6 shows the final best fit for the SED plotted in black while the intrinsic model without dust extinction is plotted in blue. The physical properties derived from the SED fit are listed in Table 3. The FTS spectrum, with the [CII] line labeled, is shown for reference and not used in the fit. The $\chi^2$ per degree of freedom is $0.82$. The importance of the results of the SED fitting and their derived properties are discussed in the next section.

\begin{table}
\caption{Photometry of HATLAS J132427} 
\begin{center} 
\begin{tabular}{c c c} 
\hline\hline\\ [0.1ex] 
Instrument & $\lambda$ &$ S_{\nu}$ \\ [0.5ex] 
\hline 
CHFT ($r$ band)  & $0.66~ \mu$m& $0.05 \pm 0.01~  \mu$Jy \\
CHFT ($z$ band)  & $0.98~ \mu$m & $0.09 \pm 0.01~ \mu$Jy \\
\emph{HST}  (F105W)   & $1.06~ \mu$m & $0.79 \pm 0.4~  \mu$Jy \\
\emph{HST}  (F160W)   & $1.54~ \mu$m & $1.81 \pm 0.6~  \mu$Jy \\
Keck ($H$ band)  & $1.63~ \mu$m& $2.41 \pm 0.8~   \mu$Jy \\
Keck ($K_{\rm s}$ band) & $2.20~ \mu$m& $3.92 \pm 0.6~   \mu$Jy \\
WISE W1          & $3.35~ \mu$m& $0.30 \pm 0.01~  $mJy \\
WISE W2          & $4.60~ \mu$m& $0.22 \pm 0.01~  $mJy\\
WISE W3         & $11.56~ \mu$m& $0.32 \pm 0.03~  $mJy\\
WISE W4          & $22.09~ \mu$m& $2.81 \pm 0.7~  $mJy\\
\textit{Herschel} (PACS) & $100~  \mu$m & $41 \pm 4~$         mJy \\
\textit{Herschel} (PACS) & $160~  \mu$m & $180 \pm 14~$mJy \\ 
\textit{Herschel} (SPIRE) & $250~  \mu$m & $347 \pm 25~$mJy \\ 
\textit{Herschel} (SPIRE) & $350~  \mu$m & $378 \pm 28~$mJy \\
\textit{Herschel} (SPIRE) & $500~  \mu$m & $268 \pm 21~$mJy \\
SCUBA2 JCMT &  $850~  \mu$m & $43   \pm 1.2~$mJy \\
SMA & $870~  \mu$m & $30.2   \pm 5.2~$mJy \\
PdBI & $2~$mm & $1.2 \pm 0.1~$mJy \\
CARMA & $3.5~$mm  & $200 \pm 170~ \mu$Jy \\
VLA & $4.3~$cm & $350 \pm 30~ \mu$Jy \\
VLA & $21~$cm & $1.95 \pm 0.24~$mJy \\[1ex] 
\hline\hline\\[0.5ex]
\end{tabular}
\end{center}
\label{table:PHOT2} 
\end{table}

\section{Discussion}

\begin{table}
\caption{SED fit and derived properties} 
\centering
\begin{tabular}{c c} 
\hline\hline\\ 
\multicolumn{2} {c} {SED fit}\\
\hline \\[0.5ex]
$f_{\mu}$(SFH/IR) &  $0.857^{+0.20}_{-0.35}$\\[2ex]
A$_{\rm V}$ &  $4.19^{+0.20}_{-0.24}$\\[2ex]
 M$_{*}$ &  $11.2^{+3.2}_{-3.8}~10^{10}$ M$_{\odot}$\\[2ex]
SFR &  $390^{+60}_{-57}~$M$_{\odot}$ yr$^{-1}$ \\[2ex]
 L$_{\rm dust}$ &  $46.8^{+5.6}_{-7.0}~10^{11}~ $L$_{\odot}$\\[2ex]
$M_{\rm dust}$ &  $13.9^{+3.0}_{-2.8}~10^8~ $M$_{\odot}$\\[2ex]
T$_{\rm dust}$ &  $33.9^{+2.1}_{-1.9}~$ K\\[2ex]
sSFR &  $30 \pm 2~ 10^{-10}$ yr$^{-1}$\\[2ex]
\hline\hline\\[0.5ex]
\multicolumn{2} {c} {Derived Properties}\\
\hline \\[0.5ex]
$\mu_{\rm dust}$&$4.9 \pm 1.8 $\\[2ex]
$\mu_{\rm stars}$&$16 \pm 4.3 $\\ [2ex]
r$_{\rm eff}$ Gas&$ 8.8 \pm 3.7~$kpc\\[2ex]
r$_{\rm eff}$ Dust&$3.2 \pm 1.2~$kpc\\[2ex]
$\Sigma_{\rm SFR}$&$12.2^{+6.8}_{-6.7}~ $ M$_{\odot}$ yr$^{-1}$ kpc$^{-2}$ \\[2ex]
$\Sigma_{\rm gas}$&$347 \pm 200~ $M$_{\odot}$ pc$^{-2}$ \\[2ex]
$M_{\rm gas}$&$8.6 \pm 3.3 \times 10^{10} \times \alpha_{\rm CO}~$ M$_{\odot}$\\[2ex]
Gas Fraction ($M_{\rm gas}$)/($M_{\rm star}+M_{\rm gas}$)&$0.43 \pm 0.13$\\[2ex]
\hline\hline\\[0.5ex]
$^1$ Based on \citet{Narayanan2012}
\end{tabular}%
\label{table:PHOT3} 
\end{table}

\begin{figure}
 \includegraphics[scale=.50]{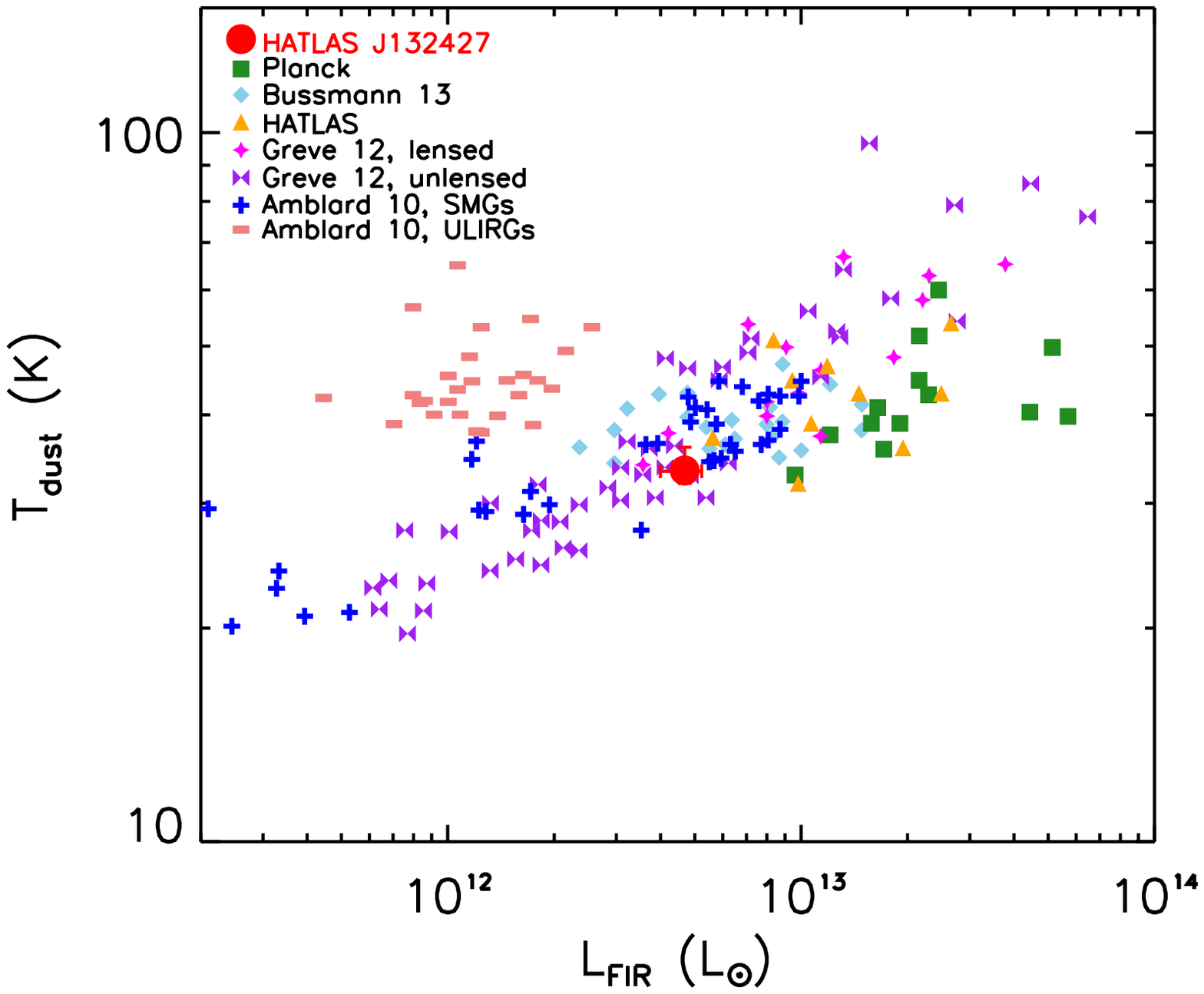}
  \caption {Dust temperature vs. FIR luminosity. For comparison other lensed and non lensed galaxies are plotted, including the other Plank/\textit{Herschel} detected lensed galaxies \citep{Canameras2015}, \textit{Herschel} lensed galaxies \citep{Bussmann2013} as well as other lensed/unlensed SMGs and ULIRGs \citep{Greve2012,Amblard2010}. To make the comparison more instructive the lensed galaxies have had L$_{\rm FIR}$ de-magnified by a factor of 5.}
\end{figure}

\begin{figure}
 \includegraphics[scale=.50]{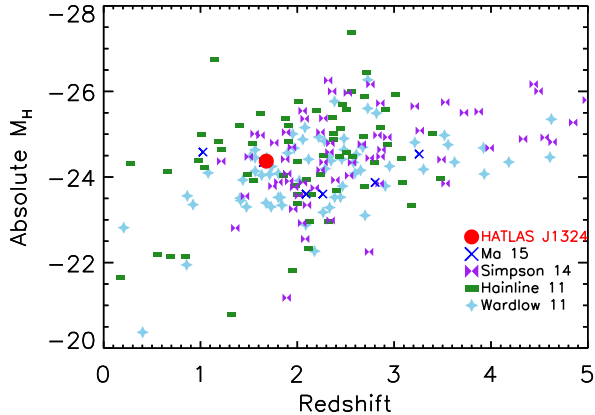}
  \caption {Rest-frame absolute $H$-band magnitude vs. redshift for DSFGs. The magnitudes have been corrected for magnification. For comparison DSFGs from other samples are included from \citet{Ma2015,Simpson2014,Hainline2011,Wardlow2013}.}
\end{figure}

\begin{figure}
\includegraphics[scale=.45]{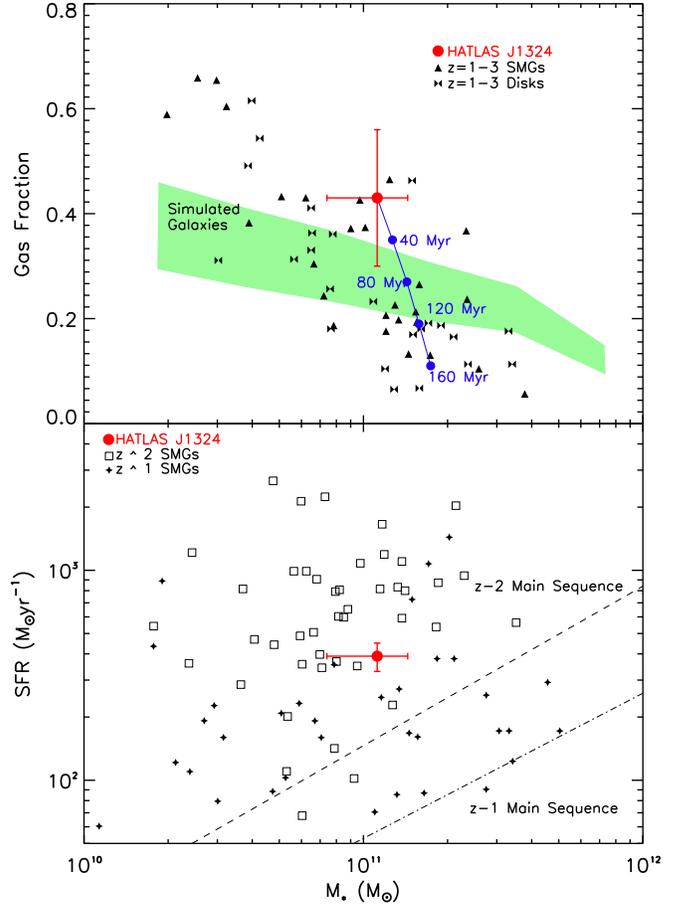}
  \caption {Top: Gas fraction vs. stellar mass. The other objects are from \citet{Narayanan2012}. The green shaded region represents star-forming galaxies at $z = 2$ from cosmological hydrodynamic simulations \citep{Dave2010}.  The blue circles represent the evolution of HATLAS J132427 over the course of 160 Myr. Each successive circle represents an 40 Myr time step with a constant SFR and mass conservation. Bottom: Star formation rate vs. stellar mass. For comparison $z \sim~2$ SMGs are plotted \citep{Fu2013} as well as $z \sim~1$ SMGs \citep{Michalowski2010,Tacconi2010,Banerji2011,Timmons2015}. The $z = 1$ and $z = 2$ main sequence \citep{Ma2015} are also plotted.}
\end{figure}

\begin{figure}
 \includegraphics[scale=.50]{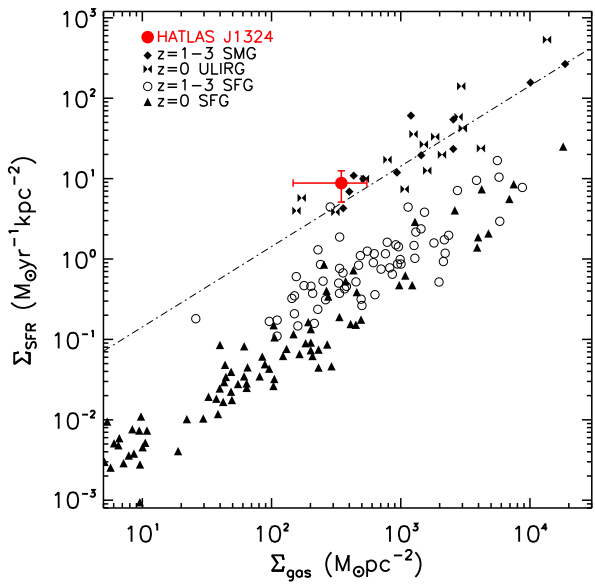}
  \caption {Star formation rate surface density vs. molecular gas surface density for local ULIRGS and SFGs as well as z $\sim$ $1-3$ SMGs and SFGs. For comparison SFGs are plotted \citep{Kennicutt1998}, as well as SMGs and local ULIRGs \citep{Tacconi2013,Fu2013}. The dashed line represents a constant gas consumption ($\tau_{\rm disk}=\Sigma_{\rm gas}/\Sigma_{\rm SFR}$) of 70 Myr for star-forming disks.}
\end{figure}

\begin{figure}
 \includegraphics[scale=.50]{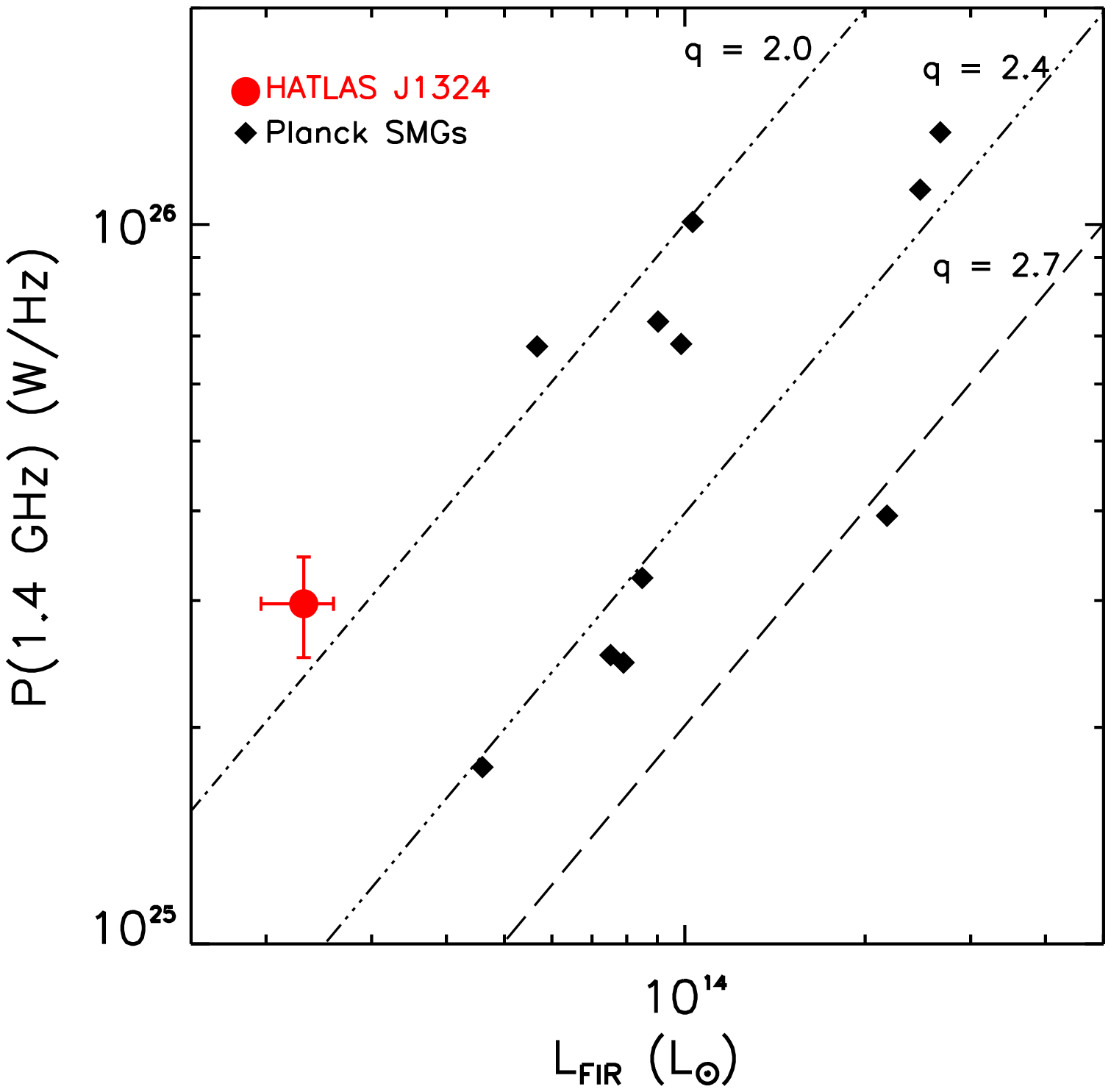}
  \caption {The far-infrared radio correlation for {\it Planck} and \textit{Herschel} detected lensed DSFGs. The other {\it Planck} detected galaxies are from \citet{Canameras2015} and represent the total number of high-redshift lensed galaxies detected in both {\it Planck} and \textit{Herschel}. The lines represent varying q values (logL$_{\rm IR}/(3.75 \times 10^{12}$~W)-logL$_{1.4}$/(W Hz$^{-1}$)) of 2.0, 2.4, and 2.7.   }
\end{figure}

Our knowledge of the physical properties of DSFGs remains limited and the goal of recent studies is to increase our understanding of the starburst phenomena in DSFGs. For this purpose we examine the various components of the galaxy, including the dust temperature, the ratio of gas to baryons, the star formation rate and its density as well as the far-infrared radio correlation. We start with a discussion of the physical properties derived from the SED fit and compare them to other SMGs and DSFGs. 

From the SED analysis, the estimated dust temperature is approximately $34~$K, which is consistent with {\it Herschel}-selected galaxies at a similar luminosity \citep{Symeonidis2013}, as well as ALESS SMGs at a similar redshift \citep{daCunha2015}. In Figure 7 we examine the relationship between dust temperature and FIR luminosity for SMGs as well as local ULIRGs. From \citet{Greve2012} the high FIR luminosity to dust temperature ratio is suggestive of a high magnification factor. In DSFGs an increased FIR luminosity correlates with an increased dust temperature. Both values come from the SED fit and are in agreement with the other high-$z$ strongly lensed galaxies. \citet{Greve2012} estimate the magnification factor for the lensed galaxies to be a factor of 1--10, which is consistent with the magnification factor ($\mu_{\rm dust} \sim 5$) from the lens model for HATLAS J132427.

Stellar masses, as derived from SED fits, depend on a few fundamental assumptions such as the assumed star formation histories (SFH), initial mass functions (IMF) and population synthesis models (see \citealt{Chabrier2003,Thomas2005,Dave2012,Michalowski2012,Conroy2013,Michalowski2014}). These introduce uncertainties in the measured stellar mass which, along with uncertainties introduced by variations in the metallicity, is usually observed as the scatter around the main sequence in the mass-SFR relation (see also \citealt{Shivaei2015} \& \citealt{Speagle2014}). Rest-frame $H$-band absolute magnitude ($M_H$) can act as a trace of stellar mass that does not depend upon an assumed SFH. In Figure 8, HATLAS J132427 is shown to have an $M_H$ consistent with other DSFGs samples and likely has a stellar mass consistent with DSFG samples.
 
To calculate the gas mass we use the CO$_{(2-1)}$ luminosity from \citet{George2013} (CO$_{(2-1)}$ = $11.3 \pm 1.4$ Jy km s$^{-1}$) and adopt a CO-H$_2$ conversion factor $\alpha_{\rm CO}$ $= 1~$M$_{\odot}$ (K km s$^{-1}~$pc$^{2}$)$^{-1}$ consistent with other SMGs (e.g., \citealt{Tacconi2008,Hodge2012}). This results in a gas mass of $8.6 \pm 3.3 \times 10^{10} \times (\alpha_{\rm CO}/1.0)$ M$_{\odot}$ assuming the magnification factor of the gas distribution to be $\mu = 4.9$, consistent with the dust. An alternative calculation for the gas mass comes from \cite{Scoville2014}, in which $28~ (z~ \textless~ 3)$ SMGs are used to find the ratio of the gas mass to the $850~ \mu$m luminosity. This ratio is found to be $1.01 \pm 0.52$. In \cite{Scoville2014} an $\alpha_{\rm CO}$ of $4.6$ is used and so here we scale the ratio down to $\alpha_{\rm CO} = 1$, giving a final ratio of L$_{850}$/$M_{\rm ISM}$ $= 0.22 \pm 0.11$. This ratio gives a gas mass of $7.3 \pm 4.6 \times 10^{10} \times (\alpha_{\rm CO}/1.0)$ M$_{\odot}$ for HATLAS J132427, which is consistent with the previous result.

The top portion of Figure 9 shows the gas-to-baryon fraction vs. stellar mass. For comparison $z =$ $1-3$ SMGs are plotted as well as z = $1-3$ main sequence star-forming galaxies. For its stellar mass, HATLAS J132427 has a large gas-to-baryon ratio ($M_{\rm gas}$)/($M_{\rm star}+M_{\rm gas}$) of $0.43$. This is in agreement with other measurements of high-$z$ star-forming galaxies \citep{Tacconi2013}. The green shaded region in Figure 9 shows star-forming galaxies at $z = 2$ from cosmological hydrodynamic simulations \citep{Dave2010}. In \citet{Narayanan2012} it is suggested that $\alpha_{\rm CO}$ is overestimated for systems at high redshift which could account for some of the scatter. The blue circles on Figure 9 represent the future evolution of HATLAS J132427 assuming a constant SFR and mass conservation. Each blue dot represents a time step 40 Myr and shows the slope of the evolution as being steeper than the overall trend of gas fraction vs. M$_*$, due to the fact that some gas must be recycled. The bottom portion of Figure 9 shows the star formation rate vs. stellar mass. Also plotted are $z \sim~1$ SMGs and $z \sim~2$ SMGs from the literature for comparison. HATLAS J132427 is above the main sequence lines for both $z = 1$ \citep{Elbaz2007} and $z = 2$ \citep{Daddi2007}. This is consistent with the large gas mass of HATLAS J132427 and its being observed in a star-bursting phase. Given the scatter in this relation, the HATLAS J132427 measured mass and star formation is different from the underlying star-forming galaxy population and is consistent with SMGs at similar redshifts.

Figure 10 shows the star formation surface density vs. molecular surface density. Plotted for comparison are $z =$ $1-3$ SMGs and SFGs, as well as local ULIRGs and SFGs. The gas area and effective gas radius is calculated using the observed gas area from CARMA and computing a de-magnified gas area based on the lens model. The dust area and effective dust radius are calculated by measuring the area of the SMA source plane reconstruction. The gas consumption time $\tau_{\rm disk}$, which refers to the star-forming disk region can be calculated using the ratio $\tau_{\rm disk} = \Sigma_{\rm gas}/\Sigma_{\rm SFR}$. For HATLAS J132427 the gas consumption time is $\sim 10~$Myr. The dot dashed line on the plot represents $\tau_{\rm disk} = 70~$Myr, which is populated with SMGs and ULIRGs while the more quiescent star-forming galaxies have $\tau_{\rm disk} \sim 1.5$ Gyr. This short timescale of star formation for HATLAS J132427 is consistent with other DSFGs.

We investigate the possibility of an AGN contribution to this source by examining the correlation between FIR and 1.4 GHz radio luminosity which is shown in Figure 11. It is common to define this correlation in terms of a value q which is defined as q = log(L$_{\rm IR}/(3.75 \times 10^{12}$~W))-log(L$_{1.4}$/(W Hz$^{-1}$)). A spectral index $\alpha = -0.8$ is assumed \citep{Condon1992}. The q value for HATLAS J132427 is $1.90$ which is lower than the average for DSFGs $\sim 2.4$ \citep{Ivison2010}. The low q value corresponds to a high relative luminosity in the radio emission and  might suggest that HATLAS J132427  has a luminous AGN (e.g.,\citealt{Vlahakis2007,Bourne2011}). It is assumed in this calculation that the radio and FIR luminosity are being magnified by the same factor. The output values of {\sc MAGPHYS} are not strongly affected by AGN contamination \citep{daCunha2015}. \citet{Hayward2015} shows that strong AGN contamination can lead to an overestimation of the stellar mass in a SED analysis.  If the longer wavelength radio is less magnified due to differential lensing the q value would be underestimated as a result.

\section{Summary}

HATLAS J132427.0+284452 is a \textit{Herschel} Astrophysical Terahertz Large Area Survey (H-ATLAS) selected strongly lensed arc of length $\sim 15''$ at $z = 1.68$. HATLAS J132427 is also {\it Planck} detected at $1.30 \pm 0.15$ Jy in the 350 $\mu$m band and is one of a few high-$z$ Planck detections in H-ATLAS. A lens model with source plane reconstructions at several wavelengths allows the estimation of magnification factors for the stars $\mu_{\rm stars} \sim 16$ and the dust $\mu_{\rm dust} \sim 5$. The different magnification values for the dust and stellar components become important for the SED analysis in which the observed fluxes must be de-magnified according to wavelength. This source demonstrates the fact that lens models constructed in a single wavelength should not be considered complete due to the effect of differential lensing.

Physical properties of the galaxy are estimated by fitting model SEDs gives a SFR of $\sim 400~ $ M$_{\odot}$ yr$^{-1}$ and a stellar mass of $\sim 11~\times 10^{10}$ M$_{\odot}$ which are consistent with a high-$z$ dusty star-forming galaxy. The SFR surface density $12~$ M$_{\odot}$ yr$^{-1}$ kpc$^{-2}$ is high compared to the molecular gas surface density $350~ $ M$_{\odot}$ pc$^{-2}$. This comes from the lens model reconstruction of the dust area which reveals a large amount of star formation is happening in a single clump. We find that the gas fraction is slightly higher than star-forming galaxies from cosmological hydrodynamic simulations but still consistent with other observations of SMGs at this redshift. The far-infrared radio correlation suggests that HATLAS J132427 might host a luminous AGN or it might be an artifact of differential lensing.

\acknowledgments
We wish to thank the anonymous referee whose comments helped improve the paper.
Financial support for this work was provided by NASA
through grant \emph{HST}-GO-13399 from the Space Telescope Science Institute, which is operated
by Associated Universities for Research in Astronomy, Inc., under NASA contract NAS 5-26555.
Additional support for NT, AC and HN was from NSF through AST-1313319.
Some of the data presented herein were obtained at the W.M. Keck Observatory, which is operated as a scientific partnership among the California Institute of Technology, the University of California and the National Aeronautics and Space Administration. The Observatory was made possible by the generous financial support of the W.M. Keck Foundation. The authors wish to recognize and acknowledge the very significant cultural role and reverence that the summit of Mauna Kea has always had within the indigenous Hawaiian community. We are most fortunate to have the opportunity to conduct observations from this mountain.
The Submillimeter Array is a joint project between the Smithsonian Astrophysical Observatory and the Academia Sinica Institute of Astronomy and Astrophysics and is funded by the Smithsonian Institution and the Academia Sinica.
C. Furlanetto acknowledges funding from CAPES (proc. 12203-1).
J.G.N. acknowledges financial support from the Spanish MINECO for a “Ramon y Cajal” fellowship.
H.M. acknowledges support from the Funda\c{c}\~ao para a Ci\^encia e a Tecnologia (FCT) through the Fellowship SFRH/BPD/97986/2013.
I.O. acknowledges support from the European Research Council in the form of the Advanced Investigator Programme, 321302, {\sc cosmicism}.
D.A.R. acknowledges support from the National Science Foundation under grant number AST-???1614213 to Cornell University.

\bibliography{NB}

\end{document}